\documentclass[runningheads]{svmult}
\usepackage{makeidx}
\usepackage{graphicx}
\usepackage{subeqnar}
\usepackage{multicol}
\usepackage{physprbb}
\makeindex 

\def\sVEV#1{\left\langle #1\right\rangle}
\def\abs#1{\left| #1\right|}
\def\braket#1{\left\langle #1\right\rangle}
\def\ket#1{\left| #1\right\rangle} 

\begin{document}
\title*{Family replicated calculation of baryogenesis\footnote{NBI-HE-02-03}}
\toctitle{8th Adriatic Meeting -- Particle Physics in the New Millennium}
\titlerunning{Family replicated calculation of baryogenesis}
\author{H.~B.~Nielsen \and Y.~Takanishi}
\authorrunning{H.~B.~Nielsen and Y.~Takanishi}
\institute{Deutsches Elektronen-Synchrotron DESY, \\
Notkestra{\ss}e 85, D-22603 Hamburg, Germany \\
\mbox{~}\\*[-0.2cm]
The Niels Bohr Institute, \\
Blegdamsvej 17, DK-2100 Copenhagen {\O}, Denmark}
\maketitle
\begin{abstract}
In our model with a Standard Model gauge group extended with 
a baryon number minus lepton number charge {\em for each
family of quarks and leptons}, we calculate the baryon number 
relative to entropy produced in early Big Bang by 
the Fukugita-Yanagida mechanism. With the parameters, 
$\hbox{\it i.e.}{}$, the Higgs 
VEVs already fitted in a very successful way to quark and lepton 
masses and mixing angles we obtain the {\em order of magnitude}
pure prediction $Y_B=2.59{+17.0\atop-2.25}\times 10^{-11}$
which according to a theoretical estimate should mean in this case 
an uncertainty of the order of a factor 7 up or down (to be 
compared to $Y_B=(1.7-8.1)\times 10^{-11}$) using a relatively crude 
approximation for the dilution factor, while using another 
estimate based 
on Buchm{\"u}ller and Pl{\"u}macher a factor $500$ less, but this 
should rather be considered a lower limit. With a realistic 
uncertainty due to wash-out of a factor $100$ up or down we 
even with the low estimate only deviate by $1.5\sigma$.
\end{abstract}
\section{Introduction}
Using the model for mass matrices presented by us in an other 
contribution~\cite{Dub} at this conference we want 
to compute the amount of baryons 
produced in the early universe. This model works by having the 
mass matrix elements being suppressed by approximately 
conserved quantum numbers from a gauge group repeated for each family 
of quarks and leptons and also having a $(B-L)$
charge for each family.

The baryon number density relative to entropy density, $Y_B$, is one 
of the rather few quantities that can give us information about the 
laws of nature beyond the Standard Model and luckily we have from 
the understanding of the production of light isotopes at 
the minute scale in Big Bang fits to this quantity~\cite{KT}. The 
``experimental'' data of the ratio of 
baryon number density to the entropy density is 
\begin{equation}
  \label{eq:YBexpnor}
  Y_B~\Big|_{\rm exp}=\left(1.7-8.1\right)\times10^{-11}\hspace{2mm}.
\end{equation}
We already had a good fit of all the masses and mixings~\cite{NT1,FNT} 
for both 
quarks and leptons measured so far and agreeing with all the bounds 
such as neutrinoless beta decay and proton decay not being seen and
matching on the borderline but consistent with the accuracy of our model 
and of the experiment of CHOOZ the electron to heaviest left-handed neutrino
mixing, and that in a version of our model in which the dominant matrix 
element in the right-handed neutrino mass matrix is the diagonal one
for the ``third'' ($\hbox{\it i.e.}{}$ with same $(B-L)_i$ as the third family) 
family $\nu_{R 3}$ right-handed neutrino.
This version of our model which fits otherwise very well does 
not give sufficient 
$(B-L)$ excess, that survives, but the by now the best model in our 
series should have the right-handed mass matrix dominated by the 
off-diagonal elements $(2,3)$ and $(3,2)$, so that there appears two
almost mass degenerate see-saw neutrinos, in addition to the 
third one (first family) which is much lighter.  
  
\section{Mass matrices and results for masses and mixing angles}
Our model produces mass matrix elements -- or 
effective Yukawa couplings  -- which are suppressed from being of the 
order of the top-mass because they are forbidden by the conservation 
of the gauge charges of our model and can only become different from 
zero using the $6$ Higgs fields~\cite{FNT,NT2} which we have in addition to 
the field replacing the Weinberg-Salam one. In the neutrino sector
according to the see-saw mechanism~\cite{seesaw} we have to 
calculate Dirac- and Majorana-mass matrices,
$M_{\rm eff} \! \approx \! M^D_\nu\,M_R^{-1}\,(M^D_\nu)^T$, to 
obtain the effective mass matrix $M_{\rm eff}$ for the 
left handed neutrinos we in practice can ``see''.
Here we present all mass matrices as they follow from our choice 
of quantum numbers for the $7$ Higgs fields in our model and for the 
quarks and leptons (as they can be found in the other contribution).
Only the quantum numbers for the field called $\phi_{B-L}$ 
is -- in order to get degenerate see-saw neutrinos -- changed into
having the $B-L$ quantum numbers of family $2$ and $3$ 
equal to $1$, $\hbox{\it i.e.}{}$, $(B-L)_2 = (B-L)_3 =1$, while the other family quantum 
numbers are just zero:

\noindent
the up-type quarks:
\begin{eqnarray}
M_{\scriptscriptstyle U} \simeq \frac{\sVEV{(\phi_{\scriptscriptstyle\rm WS})^\dagger}}{\sqrt{2}}
\hspace{-0.1cm}
\left(\!\begin{array}{ccc}
        (\omega^\dagger)^3 W^\dagger T^2
        & \omega \rho^\dagger W^\dagger T^2
        & \omega \rho^\dagger (W^\dagger)^2 T\\
        (\omega^\dagger)^4 \rho W^\dagger T^2
        &  W^\dagger T^2
        & (W^\dagger)^2 T\\
        (\omega^\dagger)^4 \rho
        & 1
        & W^\dagger T^\dagger
\end{array} \!\right)\label{M_U}
\end{eqnarray}  
\noindent
the down-type quarks:
\begin{eqnarray}
M_{\scriptscriptstyle D} \simeq \frac{\sVEV{\phi_{\scriptscriptstyle\rm WS}}}
{\sqrt{2}}\hspace{-0.1cm}
\left (\!\begin{array}{ccc}
        \omega^3 W (T^\dagger)^2
      & \omega \rho^\dagger W (T^\dagger)^2
      & \omega \rho^\dagger T^3 \\
        \omega^2 \rho W (T^\dagger)^2
      & W (T^\dagger)^2
      & T^3 \\
        \omega^2 \rho W^2 (T^\dagger)^4
      & W^2 (T^\dagger)^4
      & W T
    \end{array} \!\right) \label{M_D}
\end{eqnarray}
\noindent %
the charged leptons:
\begin{eqnarray}        
M_{\scriptscriptstyle E} \simeq \frac{\sVEV{\phi_{\scriptscriptstyle\rm WS}}}
{\sqrt{2}}\hspace{-0.1cm}
\left(\hspace{-0.1 cm}\begin{array}{ccc}
    \omega^3 W (T^\dagger)^2
  & (\omega^\dagger)^3 \rho^3 W (T^\dagger)^2 
  & (\omega^\dagger)^3 \rho^3 W T^4 \chi \\
    \omega^6 (\rho^\dagger)^3  W (T^\dagger)^2 
  &   W (T^\dagger)^2 
  &  W T^4 \chi\\
    \omega^6 (\rho^\dagger)^3  (W^\dagger)^2 T^4 
  & (W^\dagger)^2 T^4
  & WT
\end{array} \hspace{-0.1cm}\right) \label{M_E}
\end{eqnarray}
\noindent
the Dirac neutrinos:
\begin{eqnarray}
M^D_\nu \simeq \frac{\sVEV{(\phi_{\scriptscriptstyle\rm WS})^\dagger}}{\sqrt{2}}
\hspace{-0.1cm}
\left(\hspace{-0.1cm}\begin{array}{ccc}
        (\omega^\dagger)^3 W^\dagger T^2
        & (\omega^\dagger)^3 \rho^3 W^\dagger T^2
        & (\omega^\dagger)^3 \rho^3 W^\dagger  T^2 \chi\\
        (\rho^\dagger)^3 W^\dagger T^2
        &  W^\dagger T^2
        & W^\dagger T^2 \chi\\
        (\rho^\dagger)^3 W^\dagger T^\dagger \chi^\dagger
        &  W^\dagger T^\dagger \chi^\dagger
        & W^\dagger T^\dagger
\end{array} \hspace{-0.1 cm}\right)\label{Mdirac}
\end{eqnarray} 
\noindent %
and the Majorana (right-handed) neutrinos:
\begin{eqnarray}    
M_R \simeq \sVEV{\phi_{\scriptscriptstyle\rm B-L}}\hspace{-0.1cm}
\left (\hspace{-0.1 cm}\begin{array}{ccc}
(\rho^\dagger)^6 \chi^\dagger
& (\rho^\dagger)^3 \chi^\dagger /2
& (\rho^\dagger)^3/2  \\
(\rho^\dagger)^3 \chi^\dagger /2
& \chi^\dagger & 1 \\
(\rho^\dagger)^3/2 & 1 & \chi
\end{array} \hspace{-0.1 cm}\right ) \label{Majorana}
\end{eqnarray}       
We shall remember that it is here understood that all the matrix 
elements are to be provided with order of unity factors which we
do not know and in practice have treated by 
inserting random order of unity factors over which we then average at the 
end (in a logarithmic way).

\section{Renormalisation group equations}\label{RGE}
The model for the Yukawa couplings we use gives, in principle, these couplings 
at the fundamental scale, taken to be the Planck scale, at first, and we then 
use the renormalisation group to run them down to the scales where they are
to be confronted with experiment. From the Planck scale down to 
the see-saw scale or rather from 
where our gauge group is broken down to $SMG\times U(1)_{B-L}$ we use
the one-loop renormalisation group running of the Yukawa coupling constant 
matrices and the gauge couplings~\cite{NT1} in GUT notation 
including the running of Dirac neutrino Yukawa coupling:
\begin{eqnarray*}
16 \pi^2 {d g_{1}\over d  t} &\!=\!& \frac{41}{10} \, g_1^3 \hspace{2mm},\hspace{2mm}\hspace{2mm}\hspace{2mm} 16 \pi^2 {d g_{2}\over d  t} \!\!=\!\! - \frac{19}{16} \, g_2^3 \hspace{2mm}, \hspace{2mm}\hspace{2mm}\hspace{2mm} 16 \pi^2 {d g_{3}\over d  t} \!\!=\!\! - 7 \, g_3^3  \hspace{2mm},\\
16\pi^2{d Y_{\scriptscriptstyle U}\over d t}&\!=\!&\frac{3}{2}\left(Y_{\scriptscriptstyle U}(Y_{\scriptscriptstyle U})^\dagger\!-\!Y_{\scriptscriptstyle D}(Y_{\scriptscriptstyle D})^\dagger\right)Y_{\scriptscriptstyle U}\!+\!\left\{Y_{\scriptscriptstyle S}\!-\!\left(\frac{17}{20}g_1^2+\frac{9}{4}g_2^2+8g_3^2\right)\right\}Y_{\scriptscriptstyle U} ,\\
16\pi^2{d Y_{\scriptscriptstyle D}\over d t}&\!=\!&\frac{3}{2}\left(Y_{\scriptscriptstyle D}(Y_{\scriptscriptstyle D})^\dagger\!-\!Y_{\scriptscriptstyle U}(Y_{\scriptscriptstyle U})^\dagger\right)Y_{\scriptscriptstyle D}\!+\!\left\{Y_{\scriptscriptstyle S}\!-\!\left(\frac{1}{4}g_1^2+\frac{9}{4}g_2^2+8g_3^2\right)\right\}Y_{\scriptscriptstyle D} ,\\
16 \pi^2 {d Y_{\scriptscriptstyle E}\over d  t} &\!=\!& \frac{3}{2}\, 
\left( Y_{\scriptscriptstyle E} (Y_{\scriptscriptstyle E})^\dagger
-  Y_{\scriptscriptstyle \nu} (Y_{\scriptscriptstyle \nu})^\dagger\right)\,Y_{\scriptscriptstyle E} 
+ \left\{\, Y_{\scriptscriptstyle S} - \left(\frac{9}{4} g_1^2 
+ \frac{9}{4} g_2^2 \right) \right\}\, Y_{\scriptscriptstyle E} \hspace{2mm},\\
16 \pi^2 {d Y_{\scriptscriptstyle \nu}\over d  t} &\!=\!& \frac{3}{2}\, 
\left( Y_{\scriptscriptstyle \nu} (Y_{\scriptscriptstyle \nu})^\dagger -  
Y_{\scriptscriptstyle E} (Y_{\scriptscriptstyle E})^\dagger\right)\,Y_{\scriptscriptstyle \nu} 
+ \left\{\, Y_{\scriptscriptstyle S} - \left(\frac{9}{20} g_1^2 
+ \frac{9}{4} g_2^2 \right) \right\}\, Y_{\scriptscriptstyle \nu} \hspace{2mm},\\ \label{YScon} 
Y_{\scriptscriptstyle S} &\!=\!& {{\rm Tr}{}}(\, 3\, Y_{\scriptscriptstyle U}^\dagger\, Y_{\scriptscriptstyle U} 
+  3\, Y_{\scriptscriptstyle D}^\dagger \,Y_{\scriptscriptstyle D} +  Y_{\scriptscriptstyle E}^\dagger\, 
Y_{\scriptscriptstyle E} +  Y_{\scriptscriptstyle \nu}^\dagger\, Y_{\scriptscriptstyle \nu}\,) \hspace{2mm},
\end{eqnarray*}
where $t=\ln\mu$ and $\mu$ is the renormalisation point. 

In order to run the renormalisation group
equations down to $1~\mbox{\rm GeV}$, we use the following initial values:
\begin{eqnarray}
U(1):\quad & g_1(M_Z) = 0.462 \hspace{2mm},\quad & g_1(M_{\rm Planck}) = 0.614\,, \\
SU(2):\quad & g_2(M_Z) = 0.651 \hspace{2mm},\quad & g_2(M_{\rm Planck}) = 0.504\,, \\
SU(3):\quad & g_3(M_Z) = 1.22  \hspace{2mm},\quad & g_3(M_{\rm Planck}) = 0.491\,.
\end{eqnarray}
We varied the $6$ free 
parameters and found the best fit, corresponding to the lowest value 
for the quantity $\mbox{\rm g.o.f.}\equiv\sum \left[\ln \left(
\frac{\sVEV{m}_{\rm pred}}{m_{\rm exp}} \right) \right]^2=3.38$, 
with the following values for the VEVs:
\begin{eqnarray} 
\label{eq:VEVS} 
&&\sVEV{\phi_{\scriptscriptstyle WS}}= 246~\mbox{\rm GeV}\hspace{2mm},  
\hspace{2mm}\sVEV{\phi_{\scriptscriptstyle B-L}}=1.23\times10^{10}~\mbox{\rm GeV}\hspace{2mm}, 
\hspace{2mm}\sVEV{\omega}=0.245\hspace{2mm},\nonumber\\
&&\hspace{2mm}\sVEV{\rho}=0.256\hspace{2mm},\hspace{2mm}\sVEV{W}=0.143\hspace{2mm},
\hspace{2mm}\sVEV{T}=0.0742\hspace{2mm},\hspace{2mm}\sVEV{\chi}=0.0408\hspace{2mm},
\end{eqnarray}
where, except for the Weinberg-Salam Higgs field and 
$\sVEV{\phi_{\scriptscriptstyle B-L}}$, the VEVs are expressed in Planck units. 
Hereby we have considered that the Weinberg-Salam Higgs field VEV is 
already fixed by the Fermi constant.
The results of the best fit, with the VEVs in eq.~(\ref{eq:VEVS}), 
are shown in Table~\ref{convbestfit}.
\begin{table}[!b]
\begin{displaymath}
\begin{array}{|c|c|c||c|c|c|}
\hline
 & {\rm Fitted} & {\rm Experimental} &  & {\rm Fitted} & {\rm Experimental} \\\hline
m_u & 5.2~\mbox{\rm MeV} & 4~\mbox{\rm MeV} & m_d & 5.0~\mbox{\rm MeV} & 9~\mbox{\rm MeV} \\
m_c & 0.70~\mbox{\rm GeV} & 1.4~\mbox{\rm GeV} & m_s & 340~\mbox{\rm MeV} & 200~\mbox{\rm MeV} \\
M_t & 208~\mbox{\rm GeV} & 180~\mbox{\rm GeV} & m_b & 7.4~\mbox{\rm GeV} & 6.3~\mbox{\rm GeV} \\
m_e & 1.1~\mbox{\rm MeV} & 0.5~\mbox{\rm MeV} & V_{us} & 0.10 & 0.22 \\
m_{\mu} & 81~\mbox{\rm MeV} & 105~\mbox{\rm MeV} & V_{cb} & 0.024 & 0.041 \\
m_{\tau} & 1.11~\mbox{\rm GeV} & 1.78~\mbox{\rm GeV} & V_{ub} & 0.0025 & 0.0035 \\ \hline
\Delta m^2_{\odot} & 9.0 \times 10^{-5}~\mbox{\rm eV}^2 &  4.5 \times 10^{-5}~\mbox{\rm eV}^2 & \Delta m^2_{\rm atm} & 1.8 \times 10^{-3}~\mbox{\rm eV}^2 &  3.0 \times 10^{-3}~\mbox{\rm eV}^2\\
\tan^2\theta_{\odot} &0.23 & 0.35 & \tan^2\theta_{\rm atm}& 0.83 & 1.0\\ \cline{4-6} 
\tan^2\theta_{\rm chooz}  & 3.3 \times 10^{-2} & \raisebox{-.6ex}{${\textstyle\stackrel{<}{\sim}}$}~2.6 \times 10^{-2} & 
\mbox{\rm g.o.f.} &  3.38 & - \\
\hline
\end{array}
\end{displaymath}
\caption{Best fit to conventional experimental data.
All masses are running
masses at $1~\mbox{\rm GeV}$ except the top quark mass which is the pole mass.
Note that we use the square roots of the neutrino data in this 
Table, as the fitted neutrino mass and mixing parameters 
$\sVEV{m}$, in our goodness of fit ($\mbox{\rm g.o.f.}$) definition.}
\label{convbestfit}
\end{table}

\section{Quantities to use for baryogenesis calculation}
Since the baryogenesis in the Fukugita-Yanagida scheme~\cite{FY} arises from
a negative excess of lepton number being converted by Sphalerons to
a positive baryon number excess partly and this negative excess comes
from the $CP$ violating decay of the see-saw neutrinos we shall 
introduce the parameters $\epsilon_i$ giving the measure of the relative 
asymmetry under $C$ or $CP$ in the decay of neutrino number $i$: 
\noindent\ Defining the measure $\epsilon_i$ for the $CP$ violation 
\begin{equation}
  \label{eq:epsilonCP}
 \epsilon_i \equiv\frac{\sum_{\alpha,\beta}\Gamma(N_{{\scriptscriptstyle R}\, i} \to \ell^\alpha\phi_{\scriptscriptstyle WS}^\beta)-\sum_{\alpha,\beta}\Gamma(N_{{\scriptscriptstyle R}\, i}\to \bar\ell^\alpha \phi_{\scriptscriptstyle WS}^{\beta \dagger})}{\sum_{\alpha,\beta}\Gamma(N_{{\scriptscriptstyle R}\, i}
\to \ell^\alpha\phi_{\scriptscriptstyle WS}^\beta) + \sum_{\alpha,\beta}\Gamma(N_{{\scriptscriptstyle R}\, i}\to\bar\ell^\alpha \phi_{\scriptscriptstyle WS}^{\beta \dagger})}\hspace{2mm}, 
\end{equation}
where $\Gamma$ are $N_{{\scriptscriptstyle R}\, i}$ decay rates (in the $N_{{\scriptscriptstyle R}\, i}$ 
rest frame), summed over the
neutral and charged leptons (and Weinberg-Salam Higgs fields) 
which appear as final states in the $N_{{\scriptscriptstyle R}\, i}$ decays 
one sees that the 
excess of leptons over anti-leptons produced in the decay
of one $N_{{\scriptscriptstyle R}\, i}$ is just $\epsilon_i$. The total 
decay rate at the tree level is given by
\begin{equation}
  \label{eq:LOCP}
  \Gamma_{N_i}=\Gamma_{N_i\ell}+\Gamma_{N_i\bar\ell}
  ={((\widetilde{M_\nu^D})^\dagger \widetilde{M_\nu^D)}_{ii}\over 
    4\pi \sVEV{\phi_{\scriptscriptstyle WS}}^2}\,M_i \hspace{2mm},
\end{equation}%
where $\widetilde{M_\nu^D}$ can be expressed through the 
unitary matrix diagonalising the right-handed neutrino 
mass matrix $V_R$:
\begin{eqnarray}
  \label{eq:tildemd}
 \widetilde{M_\nu^D} \!&\equiv&\! M_\nu^D\,V_R \hspace{2mm},\\
V_R^\dagger \,M_R\,M_R^\dagger\, V_R \!&=&\! 
{\rm diag} \left(\,M^2_1, M^2_2, M^2_3\,\right) \hspace{2mm}.
\end{eqnarray}
The $CP$ violation rateis computed according to~\cite{CRV,BuPlu}
\begin{equation}
\label{eq:CPepsilon}
\epsilon_i = \frac{\sum_{j\not= i} {\rm Im}[((\widetilde{M_\nu^D})^{\dagger} \widetilde{M_\nu^D})^2_{ji}] \left[\, f \left( \frac{M_j^2}{M_i^2} \right) + g \left( \frac{M_j^2}{M_i^2} \right)\,\right]}{4 \pi \sVEV{\phi_{\scriptscriptstyle WS}}^2 ((\widetilde{M_\nu^D})^{\dagger}\widetilde{M_\nu^D})_{ii}}
\end{equation}
where the function, $f(x)$, comes from the one-loop vertex contribution and
the other function, $g(x)$, comes from the self-energy contribution.
These $\epsilon$'s can be calculated in perturbation theory 
only for differences between Majorana neutrino masses which 
are sufficiently large compare to their decay widths, $\hbox{\it i.e.}{}$, the 
mass splittings satisfy the condition, 
$\abs{M_i-M_j}\gg\abs{\Gamma_i-\Gamma_j}$:
\begin{equation}
f(x)=\sqrt{x} \left[1-(1+x) \ln \frac{1+x}{x}\right]\hspace{2mm}, 
\hspace{2mm} \hspace{2mm} g(x)=\frac{\sqrt{x}}{1-x} \hspace{2mm}.
\end{equation}
We as usual~\cite{KT} introduce the dacay rate relative to 
\begin{equation}
  \label{eq:Kdrei}
K_i\equiv\frac{\Gamma_i}{2 H} \,\Big|_{ T=M_{i} } = \frac{M_{\rm
Planck}}{1.66 \sVEV{\phi_{\scriptscriptstyle WS}}^2  8 \pi 
g_{*\,i}^{1/2}}\frac{((\widetilde{M_\nu^D})^{\dagger} 
\widetilde{M_\nu^D})_{ii}}{M_{i}} \qquad
(i=1, 2, 3)\hspace{2mm}, \end{equation}%
where $\Gamma_i$ is the width of the flavour $i$ Majorana neutrino,
$M_i$ is its mass and $g_{*\,i}$ is the number of degrees of freedom
at the temperature $M_i$ (in our model $\sim100$).

In order to estimate the effective $K$ factors we first 
introduce some normalized state vectors for the decay products:
\begin{equation}
\ket{ i }\equiv \left(\sum_{k=1}^{3} 
\abs{\,\left[\widetilde{M_\nu^D}(M_i)\right]_{k i}}^2\right)^{-\frac{1}{2}}\left(\,\left[ \widetilde{M_\nu^D}(M_i)\right]_{1 i} 
\,, \left[\widetilde{M_\nu^D}(M_i)\right]_{2 i}
\,, \left[\widetilde{M_\nu^D}(M_i)\right]_{3 i}\,\right)\nonumber\hspace{2mm},
\end{equation}%
\noindent
Then we may take an approximation for the effective $K$ factors:
\begin{eqnarray}
\label{eq:keff1}
{K_{\scriptscriptstyle\rm eff}}_1 &=& K_1(M_1)\hspace{2mm}, \\
\label{eq:keff2}
{K_{\scriptscriptstyle\rm eff}}_2 &=& K_2(M_2) + \abs{\braket{2 | 3}}^2 \,K_3(M_3) 
+ \abs{\braket{2 | 1}}^2\,K_1(M_1)\hspace{2mm},\\ 
\label{eq:keff3}
{K_{\scriptscriptstyle\rm eff}}_3 &=& K_3(M_3) + \abs{\braket{3 | 2}}^2\,K_2(M_2) + 
\abs{\braket{3 | 1}}^2\,K_1(M_1) \hspace{2mm}.
\end{eqnarray}

\section{Result for baryogenesis}
\indent\ Using  the Yukawa couplings -- as coming from the 
VEVs of our seven different
Higgs fields --  the numerical calculation of baryogenesis were 
performed using our random order unity factor method. In order 
to get baryogenesis in Fukugita-Yanagida scheme, we calculated the 
see-saw neutrino masses, ${K_{\scriptscriptstyle\rm eff}}_i$ 
factors and $CP$ violation parameters 
using $N=10,000$ random number 
combinations and logarithmic average method:
\begin{eqnarray*}
\label{eq:Metc} 
\begin{array}{l l l}
  M_1= 2.1 \times 10^{5}~\mbox{\rm GeV}\hspace{2mm}, & {K_{\scriptscriptstyle\rm eff}}_1= 31.6 \hspace{2mm}, & \abs{\epsilon_1} = 4.62\times 10^{-12}\hspace{2mm}\\*[0.2cm]
  M_2= 8.8 \times 10^{9}~\mbox{\rm GeV}\hspace{2mm}, & {K_{\scriptscriptstyle\rm eff}}_2 = 116.2\hspace{2mm},& 
\abs{\epsilon_2} = 4.00\times 10^{-6}\hspace{2mm} \\*[0.2cm]
  M_3= 9.9 \times 10^{9}~\mbox{\rm GeV}\hspace{2mm}, & {K_{\scriptscriptstyle\rm eff}}_3= 114.7\hspace{2mm}, & \abs{\epsilon_3} = 3.27\times 10^{-6}\hspace{2mm}
\end{array}
\end{eqnarray*}
The sign of $\epsilon_i$ is unpredictable due to the complex 
random number coefficients in mass matrices, therefore we are 
not in the position to say the sign of $\epsilon$'s. 
 Using the complex order unity random 
numbers being given by a Gaussian distribution we get after logarithmic
averaging using the dilution factors as presented by~\cite{KT,NT1}
\begin{equation}
  \label{eq:YB}
  Y_B = 2.59{+17.0\atop-2.25}\times 10^{-11} \hspace{2mm},
\end{equation}
where we have estimated the uncertainty  
in the natural exponent according ref.~\cite{FF} 
to be $64~\%\cdot\sqrt{10}\approx 200~\%$. 
 
The understanding of how this baryon to entropy prediction $Y_B$ comes
about in the model may be seen from the following (analytical) estimate
\begin{equation}
Y_B \approx \frac{1}{3}\cdot \frac{\chi}{\sqrt{g_{*}} \,T^2}\cdot\frac{M_3}{M_{\rm Planck} } \approx \frac{1}{3} \cdot 10^{-9} 
\label{analyticba}
\end{equation}
where we left out for simplicity the $\ln K$ factor in the denominator of the 
dilution factor $\kappa$ and where $M_3$ is the mass of one of the 
heavy right-handed neutrinos 
in our model $M_3\approx \sVEV{\phi_{B-L}}$. Since the atmospheric mass 
square difference square root $\sqrt{\Delta m^2_{\rm atm}}
\approx 0.05~\mbox{\rm eV} \approx \sVEV{\phi_{WS}}^2 (WT)^2/M_3$ 
we see that keeping it leaves us with the dependence 
\begin{equation}
Y_B \approx\frac{\sVEV{\phi_{WS}}^2 \chi}{3 \sqrt{0.05~\mbox{\rm eV} 
\cdot \,g_* \,M_{\rm Planck}\, W^2 T^4}}\approx \frac{1}{5}\times 10^{-4} 
\cdot \frac{\chi}{\sqrt{g_*} \,W^2 T^4} 
\end{equation}

\section{Problem with wash-out effects?}
To make a better estimate of the wash-out effect we may make 
use of the calculations by~\cite{BP} by putting effective values
for the see-saw neutrino mass $M$ and $\widetilde{m}$. The most important 
wash-out is due to ``on-shell'' formation of right-handed neutrinos 
and only depends on $K$ or the thereto proportional $\widetilde{m}$, but 
next there are wash-out effects going rather than by $K$ or $\widetilde{m}$ as 
$M \widetilde{m}^2$. In the presentation of the results by \cite{BP} fixed 
ratios between right-handed neutrino masses were assumed. However, in reality
a very important wash-out comes form the off-shell inverse decay and that 
goes as 
\begin{equation}
  \label{eq:til}
  M_1 \sum_j \frac{M_j^2}{M_1^2}\,\widetilde{m}_j^2 \hspace{2mm}\hspace{2mm} {\rm with} \hspace{2mm}\hspace{2mm}
\widetilde{m}_j \equiv
\frac{[(\widetilde{M_\nu^D})^\dagger
\widetilde{M_\nu^D}]_{jj}}{M_j}
\end{equation}
Here we use the notation with $\widetilde{m}_j$ from~\cite{BP}:
$ \widetilde{m}_j \approx K_j \cdot 2.2 \cdot 10^{-3} \mbox{\rm eV}$.

Using such a term (see eq.~\ref{eq:til}) with the ansatz 
ratios used in~\cite{BP},
$M_3^2 =10^{6}~M_1^2$ and $M_2^2 = 10^3~M_1^2$ one gets for 
eq.~(\ref{eq:til})
$\approx 10^6~M_1 ~\widetilde{m}_3^2$, while we would with our mass ratios 
(eq.~\ref{eq:Metc}) $M_3^2 \approx 1/4 \cdot 10^{10} ~M_1^2$ and 
$M_2^2 \approx 1/4 \cdot 10^{10} ~M_1^2$ obtain 
correspondingly $2\cdot 10^5~\mbox{\rm GeV} \cdot 1/4 \cdot 10^{10}
~\widetilde{m}_3^2 \approx 1/2 \cdot 10^{15}~\mbox{\rm GeV}~\widetilde{m}_3^2$, which 
then being identified with $ 10^6~M_{1 \,{\rm use}}~\widetilde{m}_3^2$  
would lead to that we should effectively use 
for simulating our model the mass of the right handed neutrino -- which
is a parameter in the presentation of the dilution effects in~\cite{BP} --
$M_{1\,{\rm use}} = 1/2 \cdot 10^{15}~\mbox{\rm GeV}/10^6 = 1/2 \cdot 
10^9~\mbox{\rm GeV}$.
Inserting this $M_{1\,{\rm use}}$ value for our estimate  $\widetilde{m}_2 
\approx \widetilde{m}_3 \approx 0.1~\mbox{\rm eV}$ gives a dilution 
factor $\kappa\approx 10^{-4}$, $\hbox{\it i.e.}{}$, a factor $500$ less 
than  what we used with our 
estimate using the $K_{{\rm eff}}$'s.
(Our $\widetilde{m}_3 =\widetilde{m}_2$ are surprisingly large compared to
the $\sqrt{\Delta m^2_{\rm atm}}$ because of renormalzation running .)
Using the better calculation of ~\cite{BP} which has a very steep dependence 
-- a fourth power say -- as function of $\widetilde{m}$ our uncertainty 
should also be corrected to a factor 100 up or down. So then we have 
one and a half standard deviations of getting too little baryon number. 

\section{Conclusion}
We calculated the baryon density relative to the entropy density 
-- baryogenesis -- from our model order of magnitudewise. This model 
already fits to quark and lepton masses and mixing 
angles using {\em only 
six parameters}, vacuum expectation values. We got a result for the 
baryon number
predicting about a factor only three less than the fitting to microwave 
background fluctuations obtained by 
Buchm{\"u}ller~$\hbox{\it et al.}$~\cite{BB}, when 
we used our crude  $K_{\rm eff}$'s approximation. However, using 
the estimate extracted from the calculations of ~\cite{BP} we got
three orders of magnitude too low prediction of the baryon number. 
This estimate must though be considered a possibly too low estimate 
because there is one scattering effect that is strongly suppressed with
our masses but which were included in that calculation. But even 
the latter estimate should because of the steep dependence of the result
on the parameters be considered more uncertain and considering the 
deviation of our prediction only $1.56\sigma$ is not unreasonable.

Since we used the Fukugita-Yanagida mechanism of obtaining first 
a lepton number excess being converted (successively by Sphalerons) 
into the baryon number, our success in this prediction should be 
considered not only a victory for our model for mass matrices 
but also for this mechanism. Since our model would be hard to 
combine with supersymmetry -- it would loose much of its 
predictive power by having to double the Higgs fields -- 
we should consider it in a \underline{non} SUSY scenario and 
thus we can without problems take the energy scale to 
inflation/reheating to be so high that the plasma had 
already had time to go roughly to thermal equilibrium 
before the right-handed neutrinos go out-of-equilibrium
due to their masses. We namely simply have no problem 
with getting too many gravitinos because gravitinos do 
not exist at all in our scheme.

Another ``unusual'' feature of our model is that the 
dominant contribution to the baryogenesis comes from 
the \underline{heavier}
right-handed neutrinos. In our model 
it could be arranged without any troubles that the two 
heaviest right-handed neutrinos have masses only deviating 
by $10\%$ namely given by our VEV parameters $\chi$. This 
leads to significant enhancement of the $\epsilon_2$ and 
$\epsilon_3$ which is crucial for the success of our prediction.
There is namely a significant wash-our taking place, 
by a factor of the order of 
$\kappa=10^{-3}~\hbox{\rm to}~10^{-6} $. It is remarkable that we have 
here worked with a model that order of magnitudewise has with only
six adjustable parameters been able to fit all the masses and mixings 
angles for quarks and leptons measured so far, including the 
Jarlskog $CP$ violation area and most importantly and 
interestingly the baryogenesis in the early Universe. To 
confirm further our model we are in strong need for further 
data -- which is not already predicted by the Standard 
Model, or we would have to improve it to give in principle 
accurate results rather than only orders of magnitudes. The 
latter would, however, be against the hall mark of our model, 
which precisely makes use of that we can guess that the huge 
amount of unknown coupling constants in our scheme with lots 
of particles can be counted as being {\em of order unity}.

\section*{Acknowledgments}
We wish to thank W.~Buchm{\"u}ller, P.~Di Bari and 
M.~Hirsch for useful discussions. We thank the Alexander 
von Humbold-Stiftung and DESY for financial support. 


\end{document}